\documentstyle[11pt]{article}
\newcommand{\bra}{\begin{array}}
\newcommand{\era}{\end{array}}
\newcommand{\beq}{\begin{equation}}
\newcommand{\eeq}{\end{equation}}
\newcommand{\beqar}{\begin{eqnarray}}
\newcommand{\eeqar}{\end{eqnarray}}

\def\BC{\bb C}
\def\_\BC{\bbi C}

\def\Tr {{\rm Tr}}

\def\( {\left(}
\def\) {\right)}
\def\[ {\left[}
\def\] {\right]}
\def\no2 {{\textstyle{n\over 2}}}

\def\Tr {{\rm Tr}}
\def\dag {{\dagger}}

\newcommand{\om}{\omega}

\newcommand{\De}{\Delta}

\newcommand{\be}{\beta}

\newcommand{\te}{\theta}

\newcommand{\al}{\alpha}

\newcommand{\lga}{\longrightarrow}

\newcommand{\da}{\dagger}

\newcommand{\lb}{\label}

\newcommand{\NP}[1]{ {\it Nucl.~Phys.} {\bf #1}}

\newcommand{\PL}[1]{ {\it Phys.~Lett.} {\bf #1}}

\newcommand{\PR}[1]{ {\it Phys.~Rev.} {\bf #1}}
\newcommand{\PRL}[1]{ {\it Phys.~Rev.~Lett.} {\bf #1}}

\newcommand{\JMP}[1]{ {\it J. Math.~Phys.} {\bf #1}}

\begin{document}
\thispagestyle{empty}
\begin{flushright}
ucd-tpg/06-03\\
hep-th/0610157
\end{flushright}

\vspace{5mm}

\begin{center}
{\Large \bf Quantum Hall Effect on the Flag Manifold ${\bf F}_2$ }

\vspace{5mm}

{\bf Mohammed Daoud}$^{a,}$\footnote{ Permanent address :
Physics Department, Faculty of Sciences, University Ibn Zohr, Agadir,
Morocco.\\ e-mail: m$_{-}$daoud@hotmail.com} and {\bf Ahmed
  Jellal}$^{b,c,}$\footnote{ e-mail : ajellal@ictp.it -- jellal@ucd.ac.ma}\\

\vspace{0.5cm}
{\em $^a$Max Planck Institute for Physics of
Complex Systems,\\ N\"othnitzer Str. 38, D-01187 Dresden,
Germany}\\

{\em $^b$Theoretical Physics Group,  Faculty of Sciences,  Choua\"ib Doukkali University,\\
Ibn M\^aachou Street,
PO Box 20,   24000 El Jadida, Morocco} \\

{\em  $^c$Center for Advanced Mathematical Sciences,
College Hall,
American University of Beirut,\\
PO Box 11-0236, Beirut, Lebanon }\\
 \vspace{3cm}

{\bf Abstract}
\end{center}

The Landau problem on the flag manifold ${\bf F}_2 =
SU(3)/U(1)\times U(1)$ is analyzed from an algebraic point of
view. The involved magnetic background is induced by two $U(1)$
abelian connections. In quantizing the theory, we show that the
wavefunctions, of a non-relativistic particle living  on  ${\bf
F}_2$,
 are the $SU(3)$ Wigner ${\cal D}$-functions
satisfying two constraints. Using the  ${\bf F}_2$ algebraic and
geometrical structures,
 we derive the Landau Hamiltonian
 as well as its energy levels.
The Lowest Landau level (LLL) wavefunctions coincide with
 the coherent states for the mixed $SU(3)$ representations.
We discuss the quantum Hall effect for a filling factor  $\nu =1$.
 where the obtained particle density is constant and
finite for a strong magnetic field. In this limit, we also show
that the system behaves like an incompressible fluid. We study
the semi-classical properties of the system confined in LLL. These
will be used to discuss the edge excitations and construct
the corresponding Wess-Zumino-Witten action.

\newpage
\section{Introduction}

Two-dimensional quantum Hall effect (QHE)~\cite{prange}
remains among the successful phenomena in condensed matter physics.
In fact, this subject continues nowadays to be investigated in different
manifolds~\cite{zhang,karabali1,karabali2,kn,knr} 
and various contexts~\cite{group0}.
The first attempt towards a high dimensional generalization of QHE was
formulated by  Hu and Zhang~\cite{zhang} on ${\bf S}^4$. Their main
motivation is based on the fact that QHE on ${\bf S}^4$ could give a way to
formulate a quantum theory of gravitation. More precisely, the edge
excitations for the quantum Hall droplet could lead to higher spin massless fields,
in particular the graviton.
Subsequently,  many
interesting studies have been done on different higher dimensional
manifolds~\cite{group0}.
Among them, Karabali and Nair~\cite{karabali1} who have employed a method
based on the group theory approach to deal with QHE and related issues on
the complex projective spaces ${\bf CP}^k$
as well as the fuzzy spaces~\cite{knr}.

The noncompact counterpart of  ${\bf CP}^k$, say the Bergman ball
${\bf B}^k$,
was considered recently both analytically \cite{jellal} and algebraically
 \cite{daoud1}. Using the group theory approach and
 considering a system of particles living on 
 ${\bf B}^k$ in the presence of a $U(1)$ background magnetic field, we have
investigated QHE. This was based on the fact that
 ${\bf B}^k$  can be viewed as the coset space  $SU(k,1)/
  U(k)$. This was used to get wavefunctions as the Wigner ${\cal D}$-functions
submitted to a set of suitable constraints and to map the
 corresponding Hamiltonian in terms of the  $SU(k,1)$ right
 generators. This latter coincides with the generalized Maass
 Laplacian in the complex coordinates. The Landau levels on  ${\bf B}^k$
 are obtained by using the correspondence between the two manifolds
${\bf CP}^k$ and
${\bf B}^k$. In
the lowest Landau levels (LLL), the obtained wavefunctions were
 nothing but
  the   $SU(k,1)$ coherent states.
Restricting to LLL, we have derived a generalized effective Wess-Zumino-Witten action that
describes the quantum Hall droplet of radius proportional to
$\sqrt{M}$, with $M$ is the number of particles in LLL. In order to
obtain the boundary excitation action, we have defined the
star product and the density of states. Also we have introduced the
perturbation potential responsible of the degeneracy lifting in terms
of the magnetic translations of $SU(k,1)$.
Finally, we have discussed the nature of the edge excitations
and illustrated this discussion by giving the disc as example.
Based on the previous results related to QHE on
${\bf CP}^k$ and ${\bf B}^k$, it is natural to consider the Landau
problem on other spaces as for instance the flag manifold  ${\bf
F}_k ={SU(k+1)/ U(1)^k}$ and discuss QHE.

The flag manifolds~\cite{picken} have appeared in physics in
different contexts as target manifolds for sigma model or in a
geometric formulation of the harmonic superspace. These  special
homogeneous spaces have interesting geometric properties, which are
relevant to discuss different issues. Indeed, they are K\"ahler
manifolds and therefore possess a symplectic form, which is
relevant to discuss QHE. This suggests to consider the Landau
problem on the  coset space ${SU(k+1)/ U(1)^k}$ and discuss its
basic features. In the present paper, we restrict ourselves to the
particular case $k=2$. The case $ k = 1$ corresponding to
two-sphere ${\bf F}_1 = {\bf CP}^1$ was considered previously in
many works, for instance see~\cite{karabali1}.

More precisely, we consider a system of particles living on the flag manifold
 ${\bf F}_2$. Taking advantage of the fact that the space  ${\bf F}_2$
can be seen as the  coset space
 ${SU(3)/ U(1)\times U(1)}$, we
analyze the quantum mechanics of the  present system. 
 Due to the geometrical nature of the considered
 manifold, we
show that the particles are submitted to the action of two magnetic
 backgrounds.
In quantizing the theory on   ${\bf F}_2$, we obtain the
wavefunctions as
 the $SU(3)$ Wigner $\cal{D}$-functions satisfying two constraints.
To derive the corresponding Hamiltonian $H$, we consider
 the right $SU(3)$ generators.
By establishing the relations between the right generators and the
 covariant derivatives, we obtain the second order differential
form of $H$.
Using the  $SU(3)$ representation theory, we
 derive the Landau energy levels indexed by four integer quantum numbers.
Restricting to
 LLL,  we find a ground state completely different
from that of the same system on  ${\bf CP}^3$ or  ${\bf R}^6$. We
analyze QHE by building the generalized Laughlin
 states
and evaluating the particle density. The incompressibility
 of these states is also considered. 
On the other hand, we analyze
the semi-classical properties of the system confined in LLL. These
will be used to discuss the edge excitations and construct
the Wess-Zumino-Witten action. 

The present paper is organized as follows. In section 2, we review
some mathematical tools related to the flag manifold ${\bf F}_2$
needed for our task. In particular we review the parametrization
of ${\bf F}_2$, mixed unitary representations and the Perelomov
coherent states of the group $SU(3)$. In section~3, by quantizing
the dynamics of a system of particles on ${\bf F}_2$, we express
the wavefunctions as the Wigner $\cal{D}$-functions satisfying two
constraints. The geometrical origin of the magnetic background
will be discussed in section 4. Also we show that the magnetic
field is a superposition of two abelian background species.
Moreover, we construct the Hamiltonian as second order
differential in terms of the ${\bf F}_2$ local coordinates. In
section 5, using the
 $SU(3)$ representation theory, we give the energy levels and wavefunctions.
We construct the Laughlin states for the fractional QHE at
$\nu={1\over m}$, with $m$ odd integer. We evaluate the particle
density as well as two-point correlation function. In fact, we
show the incompressibility of Hall system for large magnetic field
strength. In section 6, 
 we analyze the semi-classical
properties of a large collection of particles confined in LLL 
for $n_1$ and $n_2$ large. In particular, we 
derive the density distribution, the symbol associated to  a
product of two operators acting on LLL (the star product) and
give the excitation potential inducing a degeneracy
lifting. These will be used to discuss the edge excitations of a
quantum Hall droplet in the Flag manifold and constructing
their Wess-Zumino-Witten action in section 7.
We conclude and give some discussions as well as
perspectives in the last section.


\section{Flag manifold ${\bf F}_2$ }

We begin by introducing
 the flag manifold ${\bf F}_2$ and
related matters. In fact,
to discuss the quantum mechanics of a particle living on ${\bf F}_2$,
 we need to consider the parametrization of the present manifold.
 Note that, ${\bf F}_2$ is a compact
K\"ahler manifold and homogeneous but nonsymmetric parametrized
by three local complex coordinates $u_{\alpha}$, with $\alpha = 1,
2, 3$. Algebraically, ${\bf F}_2$ can be realized as the coset
space
 \beq\lb{real} {\bf F}_2 = {SU(3)/ U(1)\times U(1)}. \eeq This
realization is interesting in sense that it will allow us to use
 the group theory approach needed  for our task.
 The flag manifold  is equipped with the hermitian
Riemannian metric
\begin{equation}
ds^2 = g_{\alpha \bar\beta} du^{\alpha}d\bar{u}^{\beta}.
\end{equation}
The corresponding K\"ahler form is
 \begin{equation}
\omega = i g_{\alpha \bar\beta} du^{\alpha}d\bar{u}^{\beta}.
\end{equation}
Since that $\om$ is closed, i.e. $d\omega = 0$, the components of
the magnetic field expressed in terms of the frame fields defined
by the metric are constants. This is interesting because it
will be used to discuss QHE on the flag manifold. The metric
elements $g_{\alpha
  \bar\beta}$, which form a positive definite matrix, can be
defined
by
\begin{equation}
g_{\alpha \bar\beta} = \frac{\partial}{\partial u^{\alpha}}
\frac{\partial}{\partial \bar{u}^{\beta}}K
\end{equation}
where $K = K(u, \bar{u})$ is the K\"ahler potential, such as
\begin{equation}\lb{kpo}
K(u, \bar{u}) = \ln \left[\Delta_1(u, \bar{u})\ \Delta_2(u,
  \bar{u})\right],
\qquad u=(u_1, u_2, u_3).
\end{equation}
The functions
$\Delta_1$ and  $\Delta_2$ are given by 
\begin{equation}
\Delta_1(u, \bar{u})= 1+\vert u_1\vert^2 + \vert u_3\vert^2,\qquad
\Delta_2(u, \bar{u}) = 1+\vert u_2\vert^2 + \vert u_3 -
u_1u_2\vert^2.
\end{equation}
It is clear that $\om$ is related to $ K(u, \bar{u})$ by
\begin{equation}\lb{stform}
\omega = i \partial \bar{\partial} K.
\end{equation}
This suggests that,
 one can decompose $\om$ into two components
\beq
\om = \om_1 + \om_2
\eeq
where   $\om_1$ and  $\om_2$ read as
\beq
\om_j = i \partial \bar{\partial}\ln \Delta_j(u, \bar{u}),  \qquad j
= 1, 2.
\eeq

With the coset space realization~(\ref{real}),
an element of the manifold ${\bf F}_2$ can be written as lower triangular matrix
in terms of the local coordinates. This is
\begin{equation}\lb{uma}
u = \pmatrix{1&0&0\cr u_1&1&0\cr u_3&u_2&1\cr}. 
\end{equation}
Note that, the  elements of the group $SU(3)$ are represented by
$3\times3$ unitary matrices with determinants equal one. Moreover,
they are generated by traceless Hermitian matrices, which are
linearly independent generators $t_a$, $a = 1,2, \cdots, 8$. These
can be mapped in terms of the Gell-Mann matrices $\lambda _a$,
such as \beq t_a = \frac{\lambda _a}{2}. \eeq They verify the
normalization conditions
\begin{equation}
{\rm{Tr}}(t_at_b) = \frac{1}{2}\delta_{ab}.
\end{equation}
In terms of the matrices  $t_a$, the Weyl generators, which are the raising and lowering
operators, can be realized as
\begin{equation}
t_{\pm 1} = t_1 \pm i t_2,\qquad
t_{\pm 2} = t_4 \pm i t_5, \qquad t_{\pm 3} = t_6 \pm i t_7.
\end{equation}
The Cartan subalgebra corresponding to $SU(3)$ is generated by the elements
\begin{equation}
h_1 = \frac{1}{2}\ {\rm{diag}}(1, -1, 0),\qquad h_2 = \frac{1}{2}\
{\rm{diag}}(0, 1, -1).
\end{equation}
 From (\ref{uma}), it is clear that  ${\bf F}_2$ can be also written as
another  coset space. This is
\begin{equation}\lb{bore}
{\bf F}_2 = SL(3, {\bf C})/B_+
\end{equation}
where $ B_+$ is the Borel subgroup of the upper triangular
matrices with determinants equal to one. This is the so-called
Iwasawa decomposition~\cite{picken}. Comparing  (\ref{bore}) with
the definition~(\ref{real}), one can see that there is  an
isomorphism: 
\beq SU(3)/U(1)\times U(1) \cong SL(3, {\bf
C})/B_+. 
\eeq 
The mapping $ SU(3)/U(1) \times U(1) \rightarrow
SL(3, {\bf C})/B_+$ is a generalization of the stereographic
projection in the $SU(2)$ case.

Note that, $u$ given by (\ref{uma}) is not necessarily, in general, 
an unitary matrix. To obtain the corresponding unitary matrix $v
\in SU(3)$, we firstly consider $u$ as element of $SL(3, {\bf
C})$. It can be expressed in terms of the column vectors
\begin{equation}
u = (c_1 , c_2 , c_3) \in SL(3,{\bf C}) = SU(3)^c
\end{equation}
given by 
\beq c_1 = (1 {\hskip 0.2cm} u_1 {\hskip 0.2cm} u_3)^t,
\qquad c_2 = (0 {\hskip 0.2cm} 1 {\hskip 0.2cm} u_2)^t, \qquad c_3
= (0 {\hskip 0.2cm} 0 {\hskip 0.2cm} 1)^t 
\eeq 
where $t$ stands
for matrix transposition. Secondly, by applying the Gramm-Schmidt
orthogonalization process, we obtain, from $(c_1 , c_2 , c_3)$, a
set of mutually orthogonal vectors $(e_1 , e_2 , e_3)$. They are
\begin{equation}
e_1 = c_1, \qquad e_2 = c_2 -
\frac{(c_2,e_1)}{(e_1,e_1)}e_1, \qquad e_3 = c_3 -
\frac{(c_3,e_2)}{(e_2,e_2)}e_2 - \frac{(c_3,e_1)}{(e_1,e_1)}e_1
\end{equation}
where the inner product is defined as usually
\beq (c_i,c_j) =
c_i^t \bar{c}_j.
\eeq
Defining the normalized vectors by
\beq v_i := e_i
/ \sqrt{(e_i,e_i)}
\eeq
we get another element in $SU(3)$
mapped in terms of the local coordinates $u_{\al}$, namely
\begin{equation}
v = (v_1 , v_2 , v_3) \in  SU(3)
\end{equation}
which verifies
 $\det v = 1$ and $v^{\da}v = 1$.
Explicitly, $v$ can be written as
\begin{equation}\lb{vma}
v = \pmatrix{\frac{1}{\sqrt{\Delta _1}}&
-\frac{\bar{u}_1+u_2\bar{u}_3}{\sqrt{\Delta _1 \Delta _2}}&
-\frac{\bar{u}_3-\bar{u}_1\bar{u}_2}{\sqrt{\Delta _2}}\cr
\frac{u_1}{\sqrt{\Delta _1}} &\frac{1+\vert u_3\vert ^2 -
u_1u_2\bar{u}_3}{\sqrt{\Delta _1 \Delta _2}}&
-\frac{\bar{u}_2}{\sqrt{\Delta _2}}\cr \frac{u_3}{\sqrt{\Delta
_1}}& \frac{u_2+u_2\vert u_1\vert ^2 - u_3\bar{u}_1}{\sqrt{\Delta
_1 \Delta _2}}& \frac{1}{\sqrt{\Delta _2}}\cr}.
\end{equation}
This form is convenient to calculate the Maurer-Cartan  one-form
and then generate the magnetic background indispensable to discuss
the Landau problem as well as QHE on the flag manifold.

At this level, it is interesting to note that there is a
one-to-one correspondence between the coset representative $u \in
SU(3)/U(1)\times U(1)$ and the coherent state representation. Our
interest  in the $SU(3)$ coherent states  is mainly motivated by
the fact that they are exactly the LLL wavefunctions of
 the quantum system living on the manifold ${\bf
F}_2 $, as we will see later. The unitary irreducible
representations (UIR)  of $SU(3)$, denoted by $J\equiv\left(p,
q\right)$, are finite dimensional and
 labeled by   two positive integers $p$ and $ q$. The dimension of the corresponding Hilbert space ${\cal
H}_{\left(p, q\right)}$ is
\begin{equation}
{\rm dim} {\cal H}_{\left(p, q\right)}
 = \frac{1}{2}(p + 1)(q + 1)(p + q + 2).
\end{equation}
The orthonormal basis of ${\cal H}_{(p,q)}$ writes as
\begin{equation}\lb{base}
|\psi\rangle^{j_1,j_2, \cdots ,j_p}_{k_1, k_2, \cdots ,k_q}\equiv
|\psi\rangle^{p_1,p_2,p_3}_{q_1, q_2,q_3}, \qquad j,k =
1,2,3
\end{equation}
where the sets of non-negative integers
$(p_1,p_2,p_3)$ and $(q_1, q_2,q_3)$ satisfy two constraints
\beq\lb{pqrel}
p_1+p_2+p_3=p,\qquad
q_1+ q_2+q_3=q.
\eeq
It is well-known that $J$  can be realized via a tensor $O$
with $p$ indices belonging to UIR $(1,0)$ and $q$ indices to
UIR $(0,1)$, which  has
$(p + 1)(q + 1)(p + q + 2)/2$ complex components $O^{j_1,j_2, \cdots ,j_p}_{k_1,
k_2, \cdots ,k_q}$. It is completely symmetric separately in the
upper and lower scripts and traceless, i.e. contraction of any
upper index with any lower one gives zero. The explicit
correspondence between the tensor components  and
the basis vectors (\ref{base}) can be found in~\cite{chaturvedi}. In
 ${\cal H}_{(p,q)}$, the highest weight vector
\begin{equation}\lb{Lambda}
\vert \lambda \rangle \equiv  \vert (p,q)\lambda \rangle =
|\psi\rangle^{p,0,0}_{0,0,q}
\end{equation}
verifies the condition
\begin{equation}\label{tiaction}
 t_{+i} \vert \lambda \rangle = 0, \qquad i = 1, 2, 3.
\end{equation}
Also it is a common  eigenvector of the Cartan subalgebra
generators of $SU(3)$
\begin{equation}\label{hiaction}
 h_1 \vert \lambda \rangle = \frac{1}{2} p \vert \lambda \rangle, \qquad
  h_2 \vert \lambda \rangle = \frac{1}{2} q \vert \lambda \rangle.
\end{equation}
As we will show next, the LLL wavefunctions of the quantum
particle on ${\bf F}_2$ coincide with the $SU(3)$ coherent states.
For this, we shall sketch some important facts about the
definition and construction of the  coherent states. To begin, we
choose the highest vector $\vert \lambda \rangle$ as a reference
state and  denote by $T$ a stationary subgroup. It is defined as a
subgroup of $SU(3)$ leaving  $\vert \lambda \rangle$ invariant up
to a phase factor, namely
\begin{equation}\label{cstate}
h \vert \lambda \rangle = \vert \lambda \rangle e^{i\psi(h)},
{\hskip 1cm} h\in T.
\end{equation}
Note that, the isotropy subgroup $T$ includes the Cartan subgroup
$U(1) \times U(1)$. As any element $g \in SU(3)$ can be uniquely
decomposed into $g = \phi h$, one can have
\begin{equation}
 g \vert \lambda \rangle = \phi \vert \lambda \rangle e^{i\psi(h)}.
\end{equation}
Thus, the coherent states can be  defined by
\begin{equation}\lb{phi}
 \vert \phi, \lambda \rangle = \phi \vert \lambda \rangle
\end{equation}
and therefore they are functions of the coset space
$ SU(3)/T$.
 The maximal stability group $T$ is $U(2)$ for the
completely symmetric representation $(p,0)$ or its adjoint
$(0,q)$. In such case, $ SU(3)/T$ is the complex
projective space ${\bf CP}^2$. For a generic representation of
type $(p \neq 0,q \neq 0)$, $T = U(1)\times U(1)$ and thus the
coset space is the flag manifold ${\bf F}_2$, which is of interest
in the present analysis.

The coset representative element $\phi$
can be identified with the unitary element $v$ (\ref{vma}). It
can be written also as
\begin{equation}\lb{vvma}
v = \pmatrix{1&0&0\cr u_1&1&0\cr u_3&u_2&1\cr}
    \pmatrix{1/\sqrt{\De}_1&0&0\cr 0&\sqrt{\De_1/\De_2}&0\cr
0&0&\sqrt{\De_2}\cr}
    \pmatrix{1&\bar w_1&\bar w_3\cr 0&1&\bar w_2\cr 0&0&1\cr}
\end{equation}
where the functions $w_i$, $i = 1, 2, 3,$ are given by \beq w_1 =
-{1\over \sqrt{\De_2}} \left(u_1+\bar u_2u_3\right), \quad w_2 =
{1\over \sqrt{\De_1}} \left[\bar u_1 u_3 - u_2(1 + |u_1|^2
  )\right], \quad
w_3 = -{\sqrt{\De_1 \over \De_2}} \left(u_3 - u_1u_2 \right). \eeq
 Furthermore, in the defining
representation, one can verify that $v$ takes another form. This is
\begin{equation}
v= \exp\left(\sum_{i=1}^{3} \tau_i^- t_{-i}\right)
\exp\left[-(\ln\De_1)h_1-(\ln\De_2)h_2\right]
\exp\left(\sum_{i=1}^{3} \tau_i^+ t_{+i}\right)
\end{equation}
which is more appropriate in constructing the required coherent
states. The parameters $\tau_i^-$ and $\tau_i^+$ read as
 \beqar
 \tau_1^- &=& u_1, \qquad \tau_2^- = u_2, \qquad \tau_3^- =
u_3-\frac{1}{2}u_1u_2,\nonumber
\\
 \tau_1^+ &=& \bar w_1, \qquad
\tau_2^+ = \bar w_2, \qquad \tau_3^+ = \bar w_3-\frac{1}{2}\bar
w_1\bar w_2. 
\eeqar
 From (\ref{phi}), we can write the coherent
states as follows
\begin{equation}
\vert u_1, u_2, u_3 , \lambda \rangle := v(u_1, u_2, u_3) \vert
\lambda \rangle.
\end{equation}
To completely determine the required states $\vert u_1, u_2, u_3 , \lambda
\rangle$, we use the highest weight conditions (\ref{tiaction})
and (\ref{hiaction}). Thus, we show that
\begin{equation}\lb{vstate}
\vert u_1, u_2, u_3 , \lambda \rangle=  {\it N}(u,\bar{u})
\exp\left(\sum_{i=1}^{3} \tau_i^- t_{-i}\right) \vert \lambda
\rangle
\end{equation}
where  the normalization constant ${\it N}(u,\bar{u}) $ is 
\begin{equation}
{\it N}(u,\bar{u}) =
\Delta_1^{-\frac{p}{2}}\Delta_2^{-\frac{q}{2}}.
\end{equation}
Note that, the explicit expression of the
coherent sates (\ref{vstate}) 
has been derived in~\cite{mathur}. This derivation is based on the
Schwinger realization of the mixed representation $(p,q)$ and the
bosonic construction of the vector basis (25).

\section{Quantization of the flag manifold}

We discuss now the quantization of a particle living on the flag
manifold ${\bf F}_2$. As it will be shown, the particle is
submitted to the action of two abelian magnetic backgrounds
$U(1)$. The wavefunctions of the present system can be obtained as
functions on $SU(3)$ with specific transformation properties under
the $U(1)\times U(1)$ subgroup. In other words, the quantum
description of a "free" particle on  ${\bf F}_2$ can be performed
by reducing the free motion on the group manifold $SU(3)$. This
reduction can be established by imposing some suitable constraints
on the $SU(3)$ wavefunctions.

The classical dynamics of a free particle on $SU(3)$ is described
by the Lagrangian
\begin{equation}
L = \frac{1}{2} \Tr \left(g^{-1}\dot {g}\right)^2
\end{equation}
where dot stands for time derivative. Quantum mechanically,
the Hilbert space ${\cal H}_{(p,q)}$ is given by the square
integrable functions on the group manifold $SU(3)$, i.e. $\cal H$
= $L^2\left(SU(3)\right)$. The wavefunctions on $SU(3)$ can be
expanded as
\begin{equation}
f(g) = \sum f^J_{n_l,n_r} {\cal D}^{J}_{n_l, n _r}(g)
\end{equation}
where ${\cal D}^{J}_{n_l, n _r}(g)$  are the Wigner ${\cal
D}$-functions, such as
\begin{equation}\lb{wdf}
{\cal D}^{J}_{n _l, n _r}(g) = \langle J ,n_l \vert g \vert J,n
_r\rangle
\end{equation}
with  $g\in SU(3)$, $J \equiv(p,q)$ and
\beq
n_l \equiv ( p_{1} ,p_{2}, p_{3}, q_{1},
q_{2},q_{3})_l,  \qquad n _r\equiv ( p_{1} ,p_{2}, p_{3},
q_{1}, q_{2},q_{3})_r
\eeq
are two sets of quantum
numbers specifying the right $R_a$ and left $L_a$ actions.
The vectors $\vert J,n_r\rangle$ and $\vert J,n_l\rangle$,
 which are nothing but the ones defined by (\ref{base}), generate, respectively, the basis of $SU(3)_R$
and $SU(3)_L$ unitary irreducible representation $J$. The right $R_a$ and left $L_a$ actions
are defined by
\begin{equation}
R_a g = g t_a, \qquad L_a g = t_a g, \qquad a = 1, 2, \cdots, 8.
\end{equation}
The Wigner ${\cal D}$-functions (\ref{wdf}) are orthogonal
and form a basis of ${\cal H}_{(p,q)}$.  The states
of a $SU(3)$ representation $J$
correspond to a tensor of the form $O$ introduced previously.
Under the action of $(g_l, g_r) \in SU(3)\times SU(3)$, (\ref{wdf}) transforms as
\beq
{\cal D}^{J}_{n _l, n _r}(g)
\lga \sum_{p,q}\overline {{\cal D}^{J}_{p, n _l}(g_l)} {\cal D}^{J}_{p , q}(g)
{\cal D}^{J}_{q , n _r}(g_r).
\eeq
This relation shows that the
quantum numbers $n_r$ and $n_l$ transform, respectively,  in the
representation $J$ and the complex conjugate
representation $\bar{J}$.  Thus, ${\cal H}_{(p,q)}$ decomposes into the
sum of irreducible representations, namely
\beq
{\cal H} \cong
{\oplus} _{J} V_{\bar{J}} \otimes V_J
\eeq
where $V_J$ and
$V_{\bar{J}}$ are, respectively, the vector spaces in which the representation $J$
and  $\bar J$ are acting. The sum is over all inequivalent unitary
irreducible representations of $SU(3)$. The basis of ${\cal
H}_{(p,q)}$, introduced in the previous section, coincides
with that associated to the space $V_J$. Then, we can set the
following identification:
\begin{equation}
{\cal D}^{(p, q)}_{n_l, n _r}(g) \lga \overline{\vert
(p, q) n_l\rangle} \otimes \vert (p, q) n_r\rangle.
\end{equation}
The quantum dynamics on ${\bf F}_2$ can be described by reducing
the free motion, or imposing constraints, on the group manifold
$SU(3)$. In this sense, following the standard procedure of
quantization on the coset spaces, the classical motion on the flag
manifold is described by the Lagrangian~\cite{mullen}
\begin{equation}
L_{{\bf F}_2} = \frac{1}{2} \Tr \left(g^{-1}\dot {g}\vert _{{\bf
F}_2}\right)^2 - \Tr\left[l \left(g^{-1}\dot {g}\right)\vert
_T\right]
\end{equation}
where the symbols $\vert _T$ and  $\vert _{{{\bf F}_2}}$ stand for
the projection to the isotropy subgroup $T= U(1)\times U(1)$ and
${\bf F}_2$ in $G = SU(3)$, respectively. The first term in the
above Lagrangian is invariant under $g \to gh$ for an element $h
\in U(1)\times U(1)$ and thus depends only on ${SU(3)/ U(1)\times
U(1)}$ coordinates. The effect of reducing the motion from $SU(3)$
to the coset space ${\bf F}_2$ are contained in the second term of
the Lagrangian and given in terms of $l$. This latter can be written as
linear combination of the Cartan generators $h_1$ and $h_2$. 
It follows that to get a
quantized theory on the flag space ${\bf F}_2$, we should quantize
the following action~\cite{mullen}
\begin{equation}
S = i \int dt\ \Tr\left(l g^{-1}\dot {g}\right)
\end{equation}
where  $l $ is a combination  of the Cartan generators, such as
\beq l= n_1 h_1 + n_2 h_2. \eeq For the $ U(1)\times U(1)$
transformations of the form $g \rightarrow gh$ with 
\beq\lb{gform}
h = \exp (i\varphi_1 h_1 + i\varphi_2h_2) 
\eeq 
the action $S$
changes by a boundary term $\left(\frac{1}{2}n_1\Delta\varphi_1 +
\frac{1}{2}n_2\Delta\varphi_2 \right)$. Thus, the equations of
motion are not affected by this gauge transformation and the
classical theory is defined on the coset space ${SU(3)/ U(1)\times
U(1) }= {\bf F}_2$. The canonical momenta associated to the
direction parametrized by the angles $\varphi_1$ and $\varphi_2$,
respectively, are given by $\frac{1}{2}n_1$ and $\frac{1}{2}n_2$.
In this case, the physical states, denoted by $\psi (g)$, in the
quantum theory should satisfy two constraints. These are
\begin{equation}\lb{raction}
R_3 \psi (g) \equiv \psi (gh_1) =\frac{1}{2}n_1 \psi (g),
\qquad \frac{1}{2}(\sqrt{3}R_8 - R_3) \psi (g) \equiv
\psi (gh_2) = \frac{1}{2} n_2 \psi (g).
\end{equation}
There is another easy way to see the latter conditions. Indeed,
under the transformation $g \rightarrow gh$, the variation of the
action is given by
\beq
\Delta S =
-\frac{1}{2}\left(n_1\Delta\varphi_1 + n_2\Delta\varphi_2\right)
\eeq
and the state $\psi (g)$ transforms as
\beq
\psi (gh) = \psi
(g) \exp \left[-i \left(\frac{1}{2}n_1\varphi_1 +
\frac{1}{2}n_2\varphi_2 \right)\right].
\eeq
Using the conditions
(\ref{raction}), one can show that the right generators satisfy
the commutation relations
\beq\lb{gstate}
[R_{-1} , R_{+1}]= -
n_1,\qquad [R_{-2} , R_{+2}]= -n_1 - n_2,\qquad [R_{-3} , R_{+3}]=
-n_2 \eeq
when they  act on the states $\psi (g)$. The right
generators, or covariant derivatives,  play the role of the
creation and annihilation operators for the harmonic oscillators.
Thus, 
the groundstate 
should be annihilated by $R_{+i}$, namely 
\begin{equation}\lb{llc}
R_{+i} \psi (g) = 0.
\end{equation}
This is the so-called polarization condition in the geometric
quantization and implies that the groundstate satisfying
(\ref{llc}) is holomorphic. Physically, it describes the LLL
condition.

The wavefunctions of a quantum theory on the Flag manifold ${\bf
F}_2$ are the Wigner ${\cal D}$-functions verifying
 (\ref{raction}). As we will see later, the
polarization condition (\ref{llc}) will lead to the LLL analysis
of the present system. Note that, the constraints (\ref{raction})
and (\ref{llc}) are exactly the defining relations for a highest
weight state~(\ref{tiaction}-\ref{hiaction}). Thus, the groundstate 
wavefunctions coincide with the $SU(3)$ coherent states for
the mixed representations.


\section{Induced magnetic background}

It is well-known that the magnetic field is an important
ingredient one should define in order to formulate QHE in any
space. Thus, it is natural to ask about this physical quantity in
the present analysis. More precisely, how to generate a magnetic
background on the flag manifold  ${\bf F}_2$. This issue will be
treated by considering the geometric features of   ${\bf
F}_2$.
%

The $SU(3)$ parametrization, introduced in the first section, will
provide us with the Maurer-Cartan one-form and the $U(1)$
connections for ${SU(3)/
  U(1)\times U(1)}$. To perform this, we identify $g\in
SU(3)$ with the element $v\in {SU(3)/ T}$ given by (\ref{vma}). It
follows that a basis of invariant one-forms is given by
\begin{equation}
g^{-1} dg = -i e^{\alpha} t_{+\alpha} - i\bar e^{\alpha}
t_{-\alpha} - i \te^j h_j, \qquad \alpha = 1, 2, \quad j = 1, 2.
\end{equation}
The elements $ e^{\alpha} \equiv e^{\alpha}_{\beta} du_{\beta}$,
with summation over repeated indices, are
\beqar e^1 &=&
-\frac{i}{\Delta _1 \sqrt{\Delta _2}}\left\{ \left[1+u_3(\bar u_3
- \bar u_1\bar u_2)\right]du_1 + \left[\bar u_2-u_1(\bar u_3 -
\bar u_1\bar u_2) \right]
du_3\right\},\nonumber \\
e^2 &=& \frac{i}{\sqrt{\Delta _1 \Delta
_2}}\left(u_2du_1 - du_3\right), \nonumber\\
e^3 &=& \frac{i}{\Delta _2 \sqrt{\Delta _1}}\left\{ -u_2(\bar u_1
+ u_2 \bar u_3)du_1 - \Delta _1 du_2 + (\bar u_1 + u_2 \bar
u_3)du_3 \right\}.
\eeqar
The $U(1)$-connections  $\te^j$ are
defined by
\beq \te^j =  i du^{\alpha}  \frac{\partial}{\partial
u_{\alpha}}\ln \Delta _j + c.c., \qquad j=1,2.
\eeq
They can be
also written as
\beq \te^j =  i \te^j_{\al}du^{\alpha}+ c.c.,
\qquad \te^j_{\al} = \frac{\partial}{\partial u_{\alpha}}\ln
\Delta _j
 \eeq
 reflecting that $ \te^j$  are related to the
K\"ahler potential~(\ref{kpo}). Actually, we have two abelian
connections $\te^1$ and $\te^2$. They correspond to the vector
potentials generating the magnetic background field, under which
the quantum particle is constrained to move in the six-dimensional
manifold ${\bf F}_2$. To make contact with previous works on QHE
in higher dimensions, the present situation should be compared
with the ${\bf CP}^3$ analysis~\cite{karabali1} where the particle
is submitted only to one $U(1)$ magnetic field.
 Note that, the symplectic two-form (\ref{stform})
  can be derived from the Maurer-Cartan one-form. Indeed,
we have
\begin{equation}
\omega = - \Tr \left[2(h_1+h_2)g^{-1} dg\wedge g^{-1} dg\right].
\end{equation}
This implies
\begin{equation}
\omega = e^1 \wedge e^1 +  2 e^2 \wedge e^2 + e^3 \wedge e^3
\end{equation}
which agrees with the $\om$ form given by (\ref{stform}).

Let us denote the elements of the
inverse of the $3\times 3$ matrix $e = (e^1, e^2, e^3)$ as $(e^{-1})_{\alpha}^{\beta}$. They are
\begin{equation}
e^{-1}= i \pmatrix{\frac{\Delta_1}{\sqrt{\Delta _2}}&
-\sqrt{\frac{\Delta _1}{\Delta _2}}(u_2-u_1(\bar{u}_3-\bar
u_1\bar{u}_2))& 0 \cr 0 & \sqrt{\frac{\Delta _2}{\Delta
_1}}(\bar{u}_1+ u_2\bar{u}_3)& \frac{\Delta_2}{\sqrt{\Delta_1}}\cr
\frac{\Delta_1}{\sqrt{\Delta_2}}u_2&
\sqrt{\frac{\Delta_1}{\Delta_2}}(1 + u_3(\bar u_3 - \bar u_1\bar
u_2))& 0\cr}.
\end{equation}
To derive the Hamiltonian describing the system under
consideration, we should define the $U(1)\times U(1)$ gauge
covariant differentials on ${\bf F}_2$. In this order, from the
Maurer-Cartan one-form, we have
\begin{equation}
g^{-1}\frac{\partial g}{\partial u_{\beta}} =
-ie^{\alpha}_{\beta}t_{+\alpha} - i \theta^j_{\beta}h_j.
\end{equation}
Using this relation, one can show that the right generators,
$R_{+\alpha} g = g t_{+\alpha}$, defined by \beq R_{\pm 1} = R_1
\pm i R_2,\qquad  R_{\pm 2} = R_4 \pm i R_5,\qquad  R_{\pm 3} =
R_6 \pm i R_7 \eeq can be written as
\begin{equation}
R_{+\alpha} = i (e^{-1})_{\alpha}^{\beta} \left[\frac{\partial
}{\partial u_{\beta}} - \frac{1}{2}(n_1\te^1_{\beta} + n_2
\te^2_{\beta})\right]
\end{equation}
where we have used the constraints (\ref{raction}). They can be
mapped in terms of the gauge field as
\begin{equation}
R_{+\alpha} = i (e^{-1})_{\alpha}^{\beta} \left[\frac{\partial
}{\partial u_{\beta}} -i  a_{\be}\right]
\end{equation}
with $ a_{\be}$ given by 
\beq a_{\be}=
-\frac{i}{2}(n_1\te^1_{\beta} + n_2 \te^2_{\beta}). 
\eeq
Similarly, one can show that 
the following relation holds
\beq R_{-\alpha} =
-\overline{R}_{+\alpha}. 
\eeq 
The gauge potential can be written
as 
\beq a= a_{\be} du^{\be} + a_{\bar{\be}} du^{\bar{\be}}=
-\frac{i}{2}(n_1\te^1 + n_2 \te^2). \eeq 
Therefore the
corresponding electromagnetic field is
\beq F = da =
-\frac{i}{2}\left(n_1d\te^1 + n_2 d\te^2\right) 
\eeq 
where $n_1$ and $n_2$
are integers in agreement with the Dirac quantization. It is
obvious that $F$ is also as a superposition of two abelian parts
$F_1$ and $F_2$.

At this stage, we have the necessary ingredients to write down the
required Hamiltonian. 
It can be mapped, in terms of the $SU(3)$ right generators, as~\cite{dolan}
\begin{equation}\lb{rham}
H =  -\frac{1}{4m}\sum_{\alpha=1}^{3} \left(R_{+\alpha}
R_{-\alpha} +R_{-\alpha} R_{+\alpha} \right).
\end{equation}
By introducing
$D_{\alpha}$ and $ D_{\bar
\alpha}$
\begin{equation}\lb{dalp}
D_{\alpha} =\frac{\partial }{\partial u_{\alpha}} - \frac{\partial
}{\partial u_{\alpha}}
\ln\left(\Delta_1^{\frac{n_1}{2}}\Delta_2^{\frac{n_2}{2}}\right),
\qquad D_{\bar \alpha} = \frac{\partial }{\partial \bar
u_{\alpha}} - \frac{\partial }{\partial \bar u_{\alpha}}
\ln\left(\Delta_1^{\frac{n_1}{2}}\Delta_2^{\frac{n_2}{2}}\right)
\end{equation}
the operator $H$ takes the form
\begin{equation}\lb{ham}
H =- \frac{1}{4m} \sum_{\beta}(e^{-1})_{\beta}^{\alpha}(\bar
e^{-1})_{\beta}^{\alpha'}\left(D_{\alpha}D_{\bar{\alpha'}}+
D_{\bar{\alpha'}}D_{\alpha}\right).
\end{equation}
The forms (\ref{rham}) and  (\ref{ham}) show that there is a
bridge between the algebraic analysis and the spectral theory. The
Hamiltonian is written in terms of the local coordinates and thus
one may analytically determine the spectrum of a particle living on
the flag manifold ${\bf F}_2$. But next, we use the $SU(3)$
representation theory to get the corresponding spectrum.

\section{Spectrum and lowest Landau levels}

At this point, it is clear that to derive the spectrum of the
present system, the $U(1)$ gauge fields, or "monopoles" labeled
by two integers $n_1$ and $n_2$, will play a crucial role. Note
that, $n_1 $ and $n_2$ are related to the third component of
isospin and the hypercharge of a $SU(3)$ irreducible
representation. To analyze the Landau problem on ${\bf F}_2$, we
adopt an approach similar to that developed in~\cite{karabali1} by
studying the Landau spectrum for a quantum particle living on the
complex projective spaces ${\bf CP}^k$. As we have noticed above,
the $SU(3)$ mixed representation $(p,q)$ can be realized via the
irreducible tensor $O^p_q \equiv O^{j_1 \cdots j_p}_{k_1 \cdots
k_q}, (j,k = 1,2,3)$. It transforms under $A \in SU(3)$ according
to the rule
\begin{equation}\label{rule}
O'^{j_1 \cdots j_p}_{k_1 \cdots k_q} = A^{j_1}_{i_1}\cdots A^{j_p}_{i_p}
                                      \overline{{A^{k_1}_{l_1}}}\cdots
\overline{{A^{k_q}_{l_q}}}
                                      O^{i_1 \cdots i_p}_{l_1 \cdots l_q}.
\end{equation}

In the presence of two abelian magnetic fields, it is convenient to label
the irreducible representation $SU(3)_R$ by $(p,q)$ satisfying the
relations (\ref{pqrel})
and corresponding to the irreducible tensor $O^{p_1 , p_2 , p_3
}_{q_1 , q_2 , q_3 }$. The wavefunctions rewrite as \beq \psi (g)
= {\cal D}^ {(p_1 + p_2 + p_3 ,  q_1 + q_2 + q_3)}_{n_l,n_r}(g) =
\langle (p,q), n_l\vert g \vert (p,q), n_r\rangle. \eeq Combining
the rule transformations~(\ref{rule}) where \beq A= \exp (+i
\varphi_1 h_1 +i \varphi_2 h_2) = {\rm
diag}(e^{+\frac{i}{2}\varphi_1},
e^{-\frac{i}{2}(\varphi_1-\varphi_2)}, e^{-\frac{i}{2}\varphi_2})
\eeq and using the constraints~(\ref{raction}), we  obtain two
conditions on the integer right quantum numbers $(p_1 , p_2 , p_3
, q_1 , q_2 , q_3 )$. These are
\begin{equation}\label{k1k2}
n_1 = (p_1 - q_1)-(p_2 - q_2),\qquad
n_2 = (p_2 - q_2)-(p_3 - q_3).
\end{equation}
The states verifying~(\ref{raction}) are now
labeled by four integers. The corresponding energy levels can be
derived from of the Hamiltonian (\ref{rham}) as
\begin{equation}
E = \frac{1}{2m} \left[C_2(p,q) - R_3^2 - R_8^2 \right]
\end{equation}
where  the quadratic Casimir $C_2(p,q)$ of the $(p,q)$ representation
is given by
\beq
C_2(p,q) = \frac{1}{2}\left[p(p+3)+q(q+3)+pq\right].
\eeq
Using~(\ref{raction})  together with the constraints~(\ref{k1k2}), one
can
write $E$ as
\begin{equation}
 E(q_1 , q_2 , p_2 , p_3)=
 \frac{1}{6m}\left[3C_2\left(n_1
+ 2p_2 + p_3 + q_1 - q_2, n_2 + q_1 + 2q_2 + p_3 - p_2\right) -
(n_1^2 + n_1n_2 + n_2^2)\right].
\end{equation}
This show that actually  the Landau levels are specified by four
quantum numbers. In particular, the lowest energy eigenstates, for
$n_1$ and $ n_2$ fixed, correspond to $q_1 = q_2 = p_2 = p_3 = 0$.
This is
\begin{equation}\lb{lll}
E_0 = \frac{1}{2m} \left ( n_1 + n_2 \right)
\end{equation}
with the degeneracy
\begin{equation}
d_0 = \frac{1}{2} ( n_1 + 1)( n_2 + 1)( n_1 + n_2 + 2)
\end{equation}
This is exactly the dimension of the $(p = n_1,q = n_2 )$ representation
 or more precisely
\beq
p_1 = n_1, \qquad p_2 = p_3 = 0, \qquad q_1 = q_2 = 0,\qquad q_3 = n_2.
\eeq
The last
 constraints arise from the polarization (or lowest Landau) condition~(\ref{llc}).
 Therefore,
from (\ref{wdf}), the wavefunctions describing a free charged particle
 living on ${\bf
F}_2$ in LLL are given by
\beq\lb{psilll}
\psi_{LLL} = \langle
(n_1,n_2) (s_1,s_2,s_3)(t_1,t_2,t_3)\vert g \vert \lambda\rangle
\eeq
where $s_j , r_j \ (j = 1, 2, 3)$ stand for the left quantum
numbers of the states, which encode the degeneracy of LLL and
satisfy the relation
\beq\lb{scons}
 s_1+ s_2+s_3=n_1,\qquad  t_1+ t_2+ t_3=n_2.
\eeq
In (\ref{psilll}), $\vert \lambda \rangle$ is the highest
weight vector for the $(n_1,n_2)$ unitary irreducible
representation. As far as the flag manifold is concerned, one can
identify the group element $g$ with $v$ given by (32).
Consequently, the action of $g$ on the state $\vert \lambda
\rangle$ gives the $SU(3)$ coherent states discussed in section 2.
Thus, the LLL wavefunctions coincide with the $SU(3)$ coherent
states associated to the mixed $(n_1,n_2)$ representation. They
are given by~\cite{mathur}
\beq\lb{lllsates} \Psi_{LLL}(u_1, u_2,
u_3) =\left[{n_1!n_2!\over s_1! s_2!s_3!  t_1!
t_2!t_3!}\right]^{1\over 2} \Delta_1^{-{n_1\over
2}}\Delta_2^{-{n_2\over 2}} u_1^{ s_1} u_3^{ s_3}
(u_3-u_1u_2)^{t_1} u_2^{ t_2}.
\eeq
It is interesting to note that
the LLL wavefunctions are in correspondence with the zero modes of
the Dirac operators on the flag manifold~\cite{dolan}. We recall
that the LLL wavefunctions for complex projective space ${\bf
CP}^k$~\cite{karabali1,kn,knr} and Bergman ball ${\bf
B}^k$~\cite{daoud1} are, respectively, given by the coherent
states of the groups $SU(k+1)$ and $SU(k,1)$ in the symmetric
representations. Usually, the Perelomov coherent states for
$SU(3)$ mixed representation are
 \beq\label{cstates} 
\vert u_1, u_2, u_3 \rangle
= \sum \psi_{LLL}(u_1, u_2, u_3) \vert (n_1,n_2)
(s_1,s_2,s_3)(t_1,t_2,t_3)\rangle 
\eeq 
where the sum runs over the
quantum numbers labeling the LLL wavefunctions. They constitute
an over-complete basis 
\beq
\int d\mu\ \vert u_1, u_2, u_3 \rangle
\langle u_1, u_2, u_3 \vert = {\bf I} 
\eeq 
where ${\bf I}$ is the
identity operator and the measure 
\beq \lb{measure} 
d\mu =
\frac{(n_1+1)(n_2+1)(n_1+n_2+1)}{\pi^3 \Delta_1^2 \Delta_2^2}
\prod_{i=1}^{3}d^2u_i 
\eeq 
is simply obtained from the
$SU(3)$ Haar measure by integrating over the angles $\varphi_1$
and $\varphi_2$, see (\ref{gform}), associated to the isotropy
group $U(1)\times U(1)$. The coherent states are not orthogonal
and the overlapping  given by 
\beq\lb{ortho} 
\langle u'_1, u'_2, u'_3\vert
u_1, u_2, u_3\rangle = \left[\frac{1+\bar u'_1u_1 + \bar
u'_3u_3}{\sqrt{\Delta_1\Delta'_1}}\right]^{n_1}\left[\frac{1+\bar
u'_2u_2 + (\bar u'_3 - \bar u'_1\bar u'_2)(u_3
-u_1u_2)}{\sqrt{\Delta_2\Delta'_2}}\right]^{n_2}
\eeq
will be
useful to deal with the incompressibility of a collection of $N$
particles living on ${\bf F}_2$.

The $N$-body wavefunctions can be
obtained as the Slater determinant
\begin{equation}\lb{Ngsb1}
\Psi_N^{(1)}= \epsilon^{i_1 \cdots i_N} \Psi_{i_1}  \Psi_{i_2}
\cdots  \Psi_{i_N}
\end{equation}
where each  $\Psi_{i_j}$ has the form given
by~(\ref{lllsates})
and $\epsilon^{i_1 \cdots i_N}$ is the fully antisymmetric tensor.
This is the first Laughlin state corresponding to the filling
factor $\nu=1$. Other similar Laughlin states can be obtained as
\begin{equation}\lb{laug}
\Psi_N^{(m)}= \left\{\epsilon^{i_1 \cdots i_N} \Psi_{i_1}
\Psi_{i_2} \cdots  \Psi_{i_N}\right\}^m
\end{equation}
where $m$ is an odd integer value.

The definition  of the filling factor 
\beq \nu={N\over N_{\phi}}
 \eeq
 where $N_{\phi}$ is the quantized
flux and also represents the degrees of the Landau level
degeneracy, tells us that the particle density
 is relevant in QHE
and it should be kept constant by varying the magnetic field. In
the first Laughlin state, i.e.  $\nu = 1$, the density is given by
\begin{equation}
\rho_{0}=  \frac{( n_1 + 1)( n_2 + 1)( n_1 + n_2 + 2)}{64\pi^3
R^6}
\end{equation}
where we have 
introduced the radius $R$ of ${\bf F}_2$, such as
\beq\lb{rfield}
Ru_{\alpha} = x_{\alpha} + i x_{\alpha+3}
\eeq
and considered the volume of the flag space as~\cite{dolan} 
\beq
\mbox{vol}\left({\bf F}_2 \right) = 32\pi^3R^6.
\eeq
The thermodynamic limit corresponds to the situation in which the
radius $R$ and the number of available LLL states are large
($R\lga \infty$, $n_1,n_2 \sim n \lga \infty$). To determine the
particle density in this limit, one may use the Dirac quantization
for the flag manifold
\beq\label{dirq} 
{\bf F}_1 \equiv {\bf CP}^1
={SU(2)\over U(1)}
\eeq 
where the total magnetic field $B$ is submitted to
the constraint
\beq\label{nmagn}
n = 2BR^2.
 \eeq
 From the above tools, the density can be
approximated as 
\beq \rho_{0} \sim \left(\frac{B}{2\pi}\right)^3
\eeq 
which is constant and has a finite value. This is exactly the
particle density on the flat geometry ${\bf R}^6$ and therefore corresponds to
the fully occupied state
 $\nu=1$. It is also interesting to note that the obtained density
 coincides with that derived in the ${\bf CP}^3$ space~\cite{kn} as
 excepted since, in the limit of large radius, the
geometry of both spaces is flat.

In QHE, the quantized plateaus come from the realization of an incompressible liquid.
 This property is important since it is related to the energy. It
means that by applying an infinitesimal pressure to an
incompressible system the volume remains unchanged~\cite{ezawa}.
This condition can be checked for our system by calculating the
two-point correlation function. This can be derived by integrating
the density $\overline{{\Psi_N^{(1)}}} \Psi_N^{(1)}$ over all
particles except two. As result, we obtain
\begin{equation}\lb{tpfb1}
I(12) \sim 1 - \vert \langle u_{11}, u_{21}, u_{31} \vert u_{12},
u_{22}, u_{32} \rangle \vert^2
\end{equation}
in terms of the kernel of two states localized  at the positions
$(u_{1s}, u_{2s}, u_{3s})$, with $s = 1, 2$. Using (\ref{scons})
together with (\ref{lllsates}), it is easy to see that, in the
limit $n_1,n_2 \lga \infty$, the correlation function $I(12)$ goes
like
\begin{equation}\lb{2tpfb1}
I(12) \sim 1 - \exp \big(- n_1(|u_{11} -
u_{12}|^2+|u_{31}-u_{32}|^2)-
n_2(|u_{21}-u_{22}|^2+|u_{31}-u_{11}u_{21}-u_{32}+u_{12}u_{22}|^2)\big).
\end{equation}
This result shows that for a large magnetic field $(n_1, n_2 \sim
n = 2BR^2)$, the ${\bf F}_2$ quantum Hall system $\nu = 1$ is
incompressible and the probability to find two particles at the
same position vanishes, as it is usual in the flat geometry.

\section{Semi-classical analysis on the lowest Landau levels}

Recall that LLL of  particles living on ${\bf F}_2$ are described by 
the $SU(3)$ coherent states~(\ref{cstates}). This
provides us with a simple way to establish a correspondence
between operators and classical functions on the phase space of
the present system for large magnetic fields. In this section,
 we investigate the semi-classical
properties of a large collection of particles confined in LLL 
for $n_1$ and $n_2$ large. In particular, we
derive the density distribution, the symbol associated to a 
product of two operators acting on LLL (the star product) and
give the excitation potential inducing a degeneracy
lifting. This will be useful in driving the edge excitations of a
quantum Hall droplet in the Flag manifold.

\subsection{Density matrix and Hall droplet}

To investigate the classical behavior of a collection of particles
in LLL, we first derive the mean value of the
density matrix corresponding to an abelian droplet configuration for
large magnetic fields.
Since the coherent states (\ref{lllsates}) (LLL eigenfunctions) are labeled
by four quantum occupation numbers, one may fill the LLL states
with a large number of particles $M = N_1+N_2+M_1+M_2$ such that
the density operator is
\begin{equation}
\rho_0 = \sum_{s_1=0}^{N_1}\sum_{s_2=0}^{N_2} \sum_{t_1 =
0}^{M_1}\sum_{t_2=0}^{M_2}\vert s_1, s_2,t_1, t_2 \rangle \langle
s_1,s_2 , t_1, t_2\vert
\end{equation}
where the states read as
 \begin{equation}
\vert s_1, s_2,t_1, t_2 \rangle \equiv \vert (n_1, n_2) (s_1,
n_1-(s_1+s_2),s_2),(t_1, t_2, n_2-(t_1+t_2)) \rangle.
\end{equation}
The mean value of the density matrix is defined by
\begin{equation}
\rho_0 (\bar{u} , u)= \langle u \vert \rho_0 \vert u \rangle
\end{equation}
with $u$ stands for the variables $(u_1, u_2, u_3)$ labeling the
$SU(3)$ coherent states. $\rho_0 (\bar{u} , u)$ is
the symbol associated with the density operator. As we are
concerned with the situation when $n_1$ and $n_2$ are large, 
we analyze the spacial shape of  $\rho_0 (\bar{u} , u)$. Thus,
using~(\ref{lllsates}-\ref{cstates}), one obtains
\begin{equation}\lb{densfo}
\rho_0(\bar u, u) = \Delta_1^{-n_1}\Delta_2^{-n_2} \sum_{s_1 =
0}^{N_1}\sum_{s_2 = 0}^{N_2}\sum_{t_1 = 0}^{M_1}\sum_{t_2 =
0}^{M_2} \frac{n_1!n_2!}{s_1!s_2!t_1!t_2!}\frac{|u_1|^{2s_1}
|u_3|^{2s_2}}{(n_1-(s_1+ s_2))!}\frac{|u_3-u_1u_2|^{2t_1}
|u_2|^{2t_2}}{(n_2-(t_1+ t_2))!}.
\end{equation}
For $n_1$ and $n_2$ large, we get
\begin{equation}\label{del12}
\Delta_1^{-n_1}\Delta_2^{-n_2} = \exp(-n_1(|u_1|^{2}+
|u_3|^{2}))\exp(-n_2(|u_3-u_1u_2|^{2}+ |u_2|^{2})).
\end{equation}
Furthermore, one can verify the relation
\begin{equation}
\sum_{s_1 = 0}^{N_1} \sum_{s_2 = 0}^{N_2} \frac{n_1!}{s_1!
s_2!}\frac{|u_1|^{2s_1}|u_3|^{2s_2}}{(n_1-(s_1+s_2))!} = \sum_{s =
0}^{N_1+N_2}\frac{( n_1(|u_1|^{2}+ |u_3|^{2}))^{s}}{s!}
\end{equation}
as well as
\begin{equation}\label{formu2}
\sum_{t_1 = 0}^{M_1} \sum_{t_2 = 0}^{M_2} \frac{n_2!}{t_1!
t_2!}\frac{|u_3-u_1u_2|^{2t_1}|u_2|^{2t_2}}{(n_2-(t_1+t_2))!} =
\sum_{t = 0}^{M_1+M_2}\frac{( n_2(|u_3 -u_1u_2|^{2}+
|u_2|^{2}))^{t}}{t!}.
\end{equation}
It follows that the term involving the sum in the expression of
$\rho_0$ behaves like
\begin{equation}\label{sum3}
\sum_{s=0}^M \frac{( n_1(|u_1|^{2}+ |u_3|^{2}) +  n_2(|u_3
-u_1u_2|^{2}+ |u_2|^{2}))^s}{s!}.
\end{equation}
Combining (\ref{del12}) and (\ref{sum3}),  the density can be
approximated by
\begin{equation}
\rho_0(\bar u, u) \simeq \Theta (M - ( n_1(|u_1|^{2}+ |u_3|^{2}) +
n_2(|u_3 -u_1u_2|^{2}+ |u_2|^{2})))
\end{equation}
for a large number $M$ of particles. Clearly, $\rho_0(\bar u, u)$
is a step function for $n_1,n_2\longrightarrow \infty$ and
$M\longrightarrow \infty$ $\left(\frac{M}{n_1}, \frac{M}{n_2}\
\mbox{fixed} \right)$. Note that, a large magnetic field corresponds to a
large radius $R$, see (\ref{rfield}) and (\ref{nmagn}), one can identify
$u_3-u_1u_2$ with $u_3$. Then, introducing the rescaled variables
\begin{equation}
z_1 = \sqrt{\frac{n_1}{n}}u_1, \qquad z_2 =
\sqrt{\frac{n_2}{n}}u_2, \qquad z_3 =
\sqrt{\frac{n_1+n_2}{n}}u_3
\end{equation}
the density function takes the simple form
\begin{equation}\label{fdense}
\rho_0(\bar z, z) \simeq \Theta (M - n\bar {z}\cdot z)
\end{equation}
where dot stands for the usual scalar product and $n$ is related
to the total magnetic field defined by (\ref{rfield}). Clearly,
(\ref{fdense}) corresponds to a droplet configuration with boundary defined
by $n\bar{z}\cdot z = M$ and its radius is proportional to $\sqrt{M}$.
The derivative of this density tends to a $\delta$-function. This property play 
a crucial role  in deriving  the edge
excitations, see next.

\subsection{Star product and Moyal bracket}

An important tool to write the action describing the edge
excitations of a quantum Hall droplet in ${\bf F}_2$ is the star product . In fact for
$n_1$ and $n_2$ large the mean value of the product of two
operators leads to the Moyal star product. To show this, to every
operator $A$ acting on LLL, we associate the function
\begin{equation}\label{afunct}
{\cal A}(\bar u, u) = \langle u | A | u \rangle = \langle u_1,
u_2, u_3 | A | u_1, u_2, u_3 \rangle.
\end{equation}
An associative star product of two functions ${\cal A}(\bar u, u)$
and ${\cal B}(\bar u, u)$ is defined by
\begin{equation}\label{sprod}
{\cal A}(\bar u, u)\star {\cal B}(\bar u, u) = \langle u | AB | u
\rangle = \int d\mu(\bar{u'}, u') \langle u | A | u'
\rangle\langle u'| B | u\rangle
\end{equation}
where the measure $d\mu(\bar u, u)$ is given by (\ref{measure}). To calculate
(\ref{sprod}), we exploit the analytical
properties of coherent states defined above.
 Indeed, using (\ref{lllsates}-\ref{cstates}), one can see
that the function 
\begin{equation}
{\cal {A}}(\bar u', u) = \frac{\langle u' | A | u \rangle}{\langle
u'  | u \rangle}
\end{equation}
satisfies the holomorphic and anti-holomorphic conditions:
\begin{equation}
\frac{\partial}{\partial \bar{u}_i}{\cal {A}}(\bar u', u) = 0, \qquad
\frac{\partial}{\partial u'_i}{\cal {A}}(\bar u', u) =
0, \qquad i = 1, 2, 3, \ \ u\neq u'.
\end{equation}
Consequently, the action of the
translation operator
 on  ${\cal {A}}(\bar u', u)$ gives
\begin{equation}\label{equ1}
\exp\left(u'.\frac{\partial}{\partial u}\right){\cal {A}}(\bar u',
u) = {\cal {A}}(\bar u', u+u').
\end{equation}
This gives  ${\cal {A}}(\bar u, u')$ in terms of the function ${\cal {A}}(\bar u, u)$, namely
\begin{equation}\label{equ2}
\exp\left(-u\cdot\frac{\partial}{\partial
u'}\right)\exp\left(u'\cdot\frac{\partial}{\partial u}\right){\cal
{A}}(\bar u, u) =
 \exp\bigg((u'-u)\cdot\frac{\partial}{\partial u}\bigg){\cal {A}}(\bar u, u) = {\cal {A}}(\bar u, u').
\end{equation}
 Similarly, one
obtains
\begin{equation}\label{eq2222}
\exp\left(-\bar u\cdot\frac{\partial}{\partial \bar
u'}\right)\exp\left(\bar u'\cdot\frac{\partial}{\partial \bar
u}\right){\cal {A}}(\bar u, u) = {\cal {A}}(\bar u', u).
\end{equation}
Equivalently, (\ref{equ2}-\ref{eq2222}) can also be cast in
the following forms
\begin{equation}
 \exp\left((u'-u)\cdot\frac{\partial}{\partial u}
\right){\cal {A}}(\bar u, u) = {\cal {A}}(\bar u, u'),
\qquad
 \exp\left((\bar u'- \bar u)\cdot\frac{\partial}
{\partial \bar u}\right){\cal {A}}(\bar u, u) = {\cal {A}}(\bar u', u).
\end{equation}
Combining all 
we write
the star product as
\begin{equation}\label{starab}
{\cal A}(\bar u, u)\star {\cal B}(\bar u, u) =  \int
d\mu(\bar{u'}, u') \exp\left((u'-u)\cdot\frac{\partial}{\partial
u}\right){\cal {A}}(\bar u, u) \vert\langle u | u'
\rangle\vert^2\exp\left((\bar u'- \bar u)\cdot\frac{\partial}{\partial
\bar u}\right){\cal {B}}(\bar u, u)
\end{equation}
where the overlapping of coherent states is given by (\ref{ortho}). 
 For large magnetic field,
it  can be expressed as
\begin{equation}\label{newortho}
\langle u \vert u'\rangle = \exp\left(n\bar z \cdot z'\right)
\exp\left(-\frac{n}{2}\bar z\cdot z\right)
\exp\left(-\frac{n}{2}\bar z'\cdot z'\right).
\end{equation}
Clearly, the modulus of the kernel (\ref{ortho}) possesses the properties
$\vert\langle u | u' \rangle\vert = 1$ if and only if $u=u'$,
$\vert\langle u | u' \rangle\vert<1$
 and $\vert\langle u | u' \rangle\vert \to 0$ for $n_1$ and $n_2$ large.
This provides us with a simple way to calculate the star product
between two functions. Indeed, one can see from (\ref{newortho})
that the quantity $\vert\langle u | u' \rangle\vert$ gives
contribution only in the domain near to point $u' \simeq u$. It
follows that the sum (\ref{starab}) can be evaluated by decomposing the
integral near this point and integrating over $\eta = u'-u$. Thus, we
get
\begin{equation}
{\cal A}(\bar u, u)\star {\cal B}(\bar u, u) =  \int
\frac{d\eta.d\bar{\eta}}{\pi^3}
\exp\left(\eta\cdot\frac{\partial}{\partial u}\right){\cal {A}}(\bar u,
u)
\exp\left(-s(\eta,\bar\eta)\right)\exp\left(\bar{\eta}\cdot\frac{\partial}{\partial
\bar u}\right){\cal {B}}(\bar u, u).
\end{equation}
where 
\beq
s(\eta,\bar\eta) =  n_1(|\eta_1|^{2}+ |\eta_3|^{2}) +
n_2(|\eta_3|^{2}+ |\eta_2|^{2}). 
\eeq
Finally, by a direct
calculation, one verifies that the star product between two
functions is
\begin{equation}
{\cal A}(\bar u, u)\star {\cal B}(\bar u, u) = {\cal A} {\cal B} -
\left(\frac{1}{n_1} \frac{\partial{\cal
A}}{\partial{u_1}}\frac{\partial{\cal B}}{\partial{\bar u_1}}+
\frac{1}{n_2} \frac{\partial{\cal
A}}{\partial{u_2}}\frac{\partial{\cal B}}{\partial{\bar u_2}}
+\frac{1}{n_1+n_2} \frac{\partial{\cal
A}}{\partial{u_3}}\frac{\partial{\cal B}}{\partial{\bar
u_3}}\right)+ O\left(\frac{1}{n^2}\right).
\end{equation}
Then, the symbol or function associated with the commutator of two
operators $A$ and $B$  
\begin{equation}
\langle u |[ A , B] | u\rangle = \{{\cal A}(\bar u, u), {\cal
B}(\bar u, u)\}_{\star} 
\end{equation}
is given in terms of the Moyal bracket
\begin{equation}
 \{{\cal A}(\bar u, u), {\cal B}(\bar u,
u)\}_{\star} = {\cal A}(\bar u, u)\star {\cal B}(\bar u, u) -
{\cal B}(\bar u, u)\star {\cal A}(\bar u, u).
\end{equation}
This will be helpful in building the WZW action describing the edge
excitations.

\subsection{Excitation potential}

Note that LLL is degenerate and the degeneracy is given
by (\ref{lll}). To generate excitations, we consider the
Hamiltonian
\begin{equation}
 H_0 = E_0 + V
\end{equation}
where $E_0$ is the LLL energy (\ref{lll}) and $V$ is the excitation
potential defined by
\begin{equation}
V \vert s_1,s_2,t_1,t_2 \rangle = \omega(s_1+s_2+t_1+t_2)\vert
s_1,s_2,t_1,t_2 \rangle.
\end{equation}
The perturbation $V$ induces a  lifting of the LLL degeneracy.
Using  (\ref{cstates}), one can show that the symbol ${\cal V}(\bar u, u)$ associated to
$V$ is
\begin{equation}
\langle u\vert V \vert u \rangle = {\cal V}(\bar u, u) = \omega(
n_1(|u_1|^{2}+ |u_3|^{2}) + n_2(|u_3 -u_1u_2|^{2}+ |u_2|^{2})).
\end{equation}
It can also be written 
as
\begin{equation}\label{mvepot}
 {\cal V} = n\omega \bar z\cdot z
\end{equation}
which is just the classical harmonic oscillator potential.

\section{Edge excitations and WZW action}

 The quantum droplet under consideration is specified
by the density matrix $\rho_0$. The excitations of this
configuration can be described by an unitary time evolution
operator $U$ which gives information concerning the dynamics of
the excitations around $\rho_0$. The excited states will be
characterized by a density operator: 
\beq
\rho = U \rho_0
U^{\dagger}.
\eeq
In this section, we derive the effective action for
excitations living on the edge of this quantum droplet. The
derivation is based on semi-classical analysis presented in the
previous section. As mentioned above, the dynamical information,
related to degrees of freedom of the edge states, is contained in
the unitary operator $U$. The corresponding action is~\cite{sakita}
\begin{equation}\label{action2}
S = \int dt\ \Tr \left( \rho_0 U^{\dag}(i\partial_t - H_0)U \right).
\end{equation}
It is compatible with the Liouville evolution equation for the
density matrix
\beq
i \frac{\partial \rho}{\partial t} = [ H_0 , \rho].
\eeq
To write down an effective action describing the edge excitations,
we evaluate the quantities occurring in (\ref{action2}) as classical
functions on the basis of the semi-classical analysis performed
above. Note that, the strategy adopted here is similar to those 
developed in references~\cite{karabali2,kn,daoud1}. Indeed, 
 we start by calculating the term $ i \int dt\ \Tr(\rho_0
U^{\dag}\partial_tU)$ with $U= e^{+i\Phi}$ and
$\Phi^{\dag} = \Phi$. A direct calculation gives
\begin{equation}
dU = \sum_{k=1}^{\infty}\frac{(i)^k}{k!}\sum_{p=0}^{k-1}\Phi^p
d\Phi \Phi^{k-1-p}.
\end{equation}
It leads 
\begin{equation}
U^{\dag}dU = i \int_0^1 d\alpha\ e^{-i\alpha\Phi}d\Phi
e^{+i\alpha\Phi}.
\end{equation}
Thus, we have
\begin{equation}
e^{-i\Phi}\partial_t e^{+i\Phi} = i \int_0^1 d\alpha e^{-i\alpha
\Phi}\partial_t\Phi e^{+i\alpha\Phi}.
\end{equation}
Using Baker-Campbell-Hausdorff formula, one can show
\begin{equation}
i \int dt\ \Tr(\rho_0 U^{\dag}\partial_tU) = \int dt\
\sum_{k=0}^{\infty} \frac{- (i)^k}{(k+1)!}\ \Tr
(\underbrace{[\Phi,\cdots[\Phi}_k,\rho_0]\cdots]\partial_t\Phi).
\end{equation}
Due to the coherent states completeness, the trace of any operator
$A$ is
\beq
\Tr A = \int d\mu(\bar u, u)\ \langle u | A | u \rangle.
\eeq
It follows that 
\begin{equation}
i \int dt\ \Tr(\rho_0 U^{\dag}\partial_tU) =\int d\mu dt\
\sum_{k=0}^{\infty} \frac{- (i)^k}{(k+1)!}
\underbrace{\{\Phi,\cdots\{\Phi}_k,\rho_0\}_{\star}\cdots\}_{\star}\star\partial_t\Phi
\end{equation}
where the star product and the Moyal bracket are those defined before. It is important to stress that
$\rho_0$ and $\Phi$  are now classical functions.
It is easy to obtain
\begin{equation}
i \int dt\ \Tr(\rho_0 U^{\dag}\partial_tU) \simeq -\frac {i}{2} \int
d\mu dt\ \{\Phi,\rho_0\}_{\star}\partial_t\Phi
\end{equation}
here we have dropped terms containing  the total time derivative
as well as those of higher orders. We show that 
the Moyal bracket 
reads as
\begin{equation}
\{\Phi , \rho_0\}_{\star} = \frac{i}{n}({\cal L}\Phi)
\frac{\partial\rho_0}{\partial (\bar z\cdot z)}
\end{equation}
where the first order differential operator is given by
\begin{equation}
{\cal L} =  i \left(z\cdot \frac{\partial}{\partial z} - \bar
z\cdot\frac{\partial}{\partial \bar z}\right).
\end{equation}
in terms of the rescaled variables $z_i$.  Since the
derivative of the density (\ref{fdense}) is a $\delta$-function with
support on the boundary $\partial {\cal D}$ of the droplet ${\cal
D}$ defined by $n\bar z\cdot z = M$, we get
\begin{equation}
i \int dt\ \Tr(\rho_0 U^{\dag}\partial_tU) \approx -\frac{1}{2}\int
d\mu dt\ \delta(M-n\bar z\cdot z)({\cal L}\Phi)( \partial_t\Phi ) =
-\frac{1}{2} \int_{\partial {\cal D}\times{\bf R}^+} dt\ ({\cal
L}\Phi)( \partial_t\Phi ).
\end{equation}

 Now we come to the valuation of the second term in the action
 (\ref{action2}). By
a straightforward calculation, we obtain
\begin{equation}\label{seterm}
\Tr(\rho_0 U^{\dag} V U) \simeq \Tr(\rho_0  V ) + i \Tr([\rho_0, V] \Phi)
+ \frac{1}{2}\Tr([\rho_0, \Phi ][V, \Phi ]).
\end{equation}
The first term in r.h.s of (\ref{seterm}) ($\Phi$-independent) does not
contain
any information about the edge excitations of the Hall droplet. Thus, it
can be ignored. The second term in r.h.s of (\ref{seterm}) rewrites as
\begin{equation}\lb{seterm2}
i \Tr([\rho_0, V] \Phi)  \approx i \int d\mu\ \{\rho_0, {\cal
V}\}_{\star} \Phi
\end{equation}
in terms of the Moyal bracket. 
It is easy to see that the star product in (\ref{seterm2}) is zero and we
have
\begin{equation} 
i \Tr([\rho_0, V] \Phi) \longrightarrow  0.
\end{equation}
The last term in r.h.s of (\ref{seterm}) gives
\begin{equation}
\frac{1}{2}\ \Tr([\rho_0, \Phi ][V, \Phi ]) \approx -\frac{1}{2n}\int
d\mu dt\  ({\cal L}\Phi) \frac{\partial\rho_0}{\partial (\bar z\cdot z)}
({\cal L}\Phi) \frac{\partial{\cal V}}{\partial (\bar z\cdot z)}.
\end{equation}
Using  (\ref{fdense}) and (\ref{mvepot}),  we find
\begin{equation}
  \int dt \ \Tr(\rho_0
U^{\dag} H U) =  \frac{\omega}{2}\int d\mu dt\ \delta(M-n\bar
z\cdot z)({\cal L}\Phi)^2.
\end{equation}
Note that we have eliminated the term containing the groundstate
energy $E_0$ which does not contribute to the edge dynamics.
Finally, combining all together to get
\begin{equation}
S \approx -\frac{1}{2}\int_{\partial {\cal D}\times {\bf R }^+}
d\mu dt\ \delta(M-n\bar z\cdot z)({\cal L}\Phi) \left((
\partial_t\Phi )+\omega ({\cal L}\Phi)\right).
\end{equation}
This action involves only the time derivative of $\Phi$ and the
tangential derivatives ${\cal L}\Phi$. It generalizes the chiral
abelian WZW theory describing a bosonized
theory of a system of a large number of fermions in two-dimensions~\cite{sakita}. It
is interesting that the obtained WZW action is similar to one
describing the edge excitations for Hall droplets in the six-dimensional 
complex projective ${\bf CP}^3$ \cite{karabali2} and in the
Bergman Ball ${\bf B}^3$ \cite{daoud1}.

\section{Conclusion and discussions}

We have analyzed, through this paper, some aspects of 
the quantum Hall
effect at the filling factor $\nu = 1$ on the flag manifold ${\bf F}_2$. 
More precisely, we have
algebraically investigated the eigenvalue problem of a collection
of $N$ non-interacting particles living on ${\bf F}_2$. We have
shown, in quantizing the theory, that the wavefunctions write as
the Wigner ${\cal D}$-functions. They satisfy two constraints
which are in correspondence with the $U(1)$ abelian gauge fields.
Obtaining the energy levels is an easy task thanks to the
$SU(3)$ representation theory. Also, we have derived the
analytical expression of the Landau Hamiltonian describing the
dynamics of a non-relativistic particle living on ${\bf F}_2$. We
have clearly established that the lowest Landau level
wavefunctions  coincide with $SU(3)$ coherent states expressed in
terms of the ${\bf F}_2$ coordinates. 
We have constructed
the Laughlin states describing the fractional quantum Hall effect
at $\nu =
\frac{1}{m}$, with $m$ odd integer. For the state $\nu = 1$, 
we have shown for large magnetic field that
the particle density is finite as well as constant and the system
behaves like an incompressible fluid.

On the other hand,
 we have analyzed the semi-classical
properties of a large collection of particles confined in LLL 
for $n_1$ and $n_2$ large. In particular, we have
derived the density distribution, the symbol associated to  a
product of two operators acting on LLL (the star product) and
given the excitation potential inducing a degeneracy
lifting. This is used to discuss the edge excitations of a
quantum Hall droplet in the Flag manifold and constructing
their Wess-Zumino-Witten action.

It is obvious, from previous analysis, that one can obtain the
Landau spectrum of a system living on ${\bf CP}^2$. This can be
performed by reducing the mixed representation $(p , q)$ to  the
completely symmetric one $(p , 0)$ or its adjoint $(0 , q)$. In
this way, one recovers the results of Karabali and
Nair~\cite{karabali1}. Furthermore, because they are six
dimensional, one can compare ${\bf F}_2$ and ${\bf CP}^3$
analysis~\cite{karabali1}. In the ${\bf CP}^3$ case,
QHE can be approached following two different ways. The first one
corresponds to the situation in which only one $U(1)$ abelian
gauge is involved. In the second situation, the particle evolves
in the $SU(2)$ magnetic background. Another interesting comparison
concerns the particle density for large magnetic field. It is
remarkable that, in the both spaces ${\bf F}_2$ and ${\bf CP}^3$,
we find the same value which coincides with one obtained for
particles living on ${\bf R}^6$.

To close this discussion, as examples of quantum systems submitted
to two magnetic fields, we quote the composite fermions and
multi-layers or a set of electrons and holes together. {Composite
fermions} are a new kind of particles which appear in condensed matter
physics to provide an explanation of the behavior of electrons
moving in a strong magnetic field $B$~\cite{cf}. Electrons
possessing $2l\Phi_0$, $l=1,2,\cdots$, flux quanta (vortices) can
be thought of being composite fermions. One of the most important
features of them is that they feel effectively the magnetic field 
\beq
\lb{cfm} 
B^*=B - 2l\Phi_0\rho
\eeq
 where $\rho$ is the electron
density. This magnetic field can be seen as a superposition of two
abelian parts.

Of course, the prolongations of the present work are numerous and
some interesting questions still open. The first point concerns
the analytical derivation of Landau spectrum (energies and
wavefunctions) by solving the Schr\"odinger equation for the
Hamiltonian (\ref{ham}). 
The second question is related to a
possible generalization of the present study to other higher
dimensional flag manifolds, i.e. $k\geq 3$.
Finally, one may ask about the topological excitations
on the flag manifold generalizing those constructed by
Haldane~\cite{haldane} on two-sphere ${\bf S}^2$. 

\section*{ Acknowledgments}

MD work's was partially done during a visit to the Max Planck
Institute for the Physics of Complex Systems. He would like to acknowledge the MPI-PKS
for the helpful atmosphere and financial support. MD would like to
thank the Abdus Salam ICTP where this work was finalized. AJ work's
was partially supported by the Arab Regional Fellows Program (ARFP).
The authors are indebted to the
referees for their constructive comments.

\end{document}